# The Large-Scale Structure of Semantic Networks:
# Statistical Analyses and a Model of Semantic Growth

**Mark Steyvers**
(msteyver@psych.stanford.edu)

**Joshua B. Tenenbaum**
(jbt@psych.stanford.edu)

Stanford University

**Abstract**

We present statistical analyses of the large-scale structure of three types of semantic networks: word associations, WordNet, and Roget's thesaurus. We show that they have a small-world structure, characterized by sparse connectivity, short average path-lengths between words, and strong local clustering. In addition, the distributions of the number of connections follow power laws that indicate a scale-free pattern of connectivity, with most nodes having relatively few connections joined together through a small number of hubs with many connections. These regularities have also been found in certain other complex natural networks, such as the world wide web, but they are not consistent with many conventional models of semantic organization, based on inheritance hierarchies, arbitrarily structured networks, or high-dimensional vector spaces. We propose that these structures reflect the mechanisms by which semantic networks grow. We describe a simple model for semantic growth, in which each new word or concept is connected to an existing network by differentiating the connectivity pattern of an existing node. This model generates appropriate small-world statistics and power-law connectivity distributions, and also suggests one possible mechanistic basis for the effects of learning history variables (age-of-acquisition, usage frequency) on behavioral performance in semantic processing tasks.

Network structures provide intuitive and useful representations for modeling semantic knowledge and inference. Within the paradigm of semantic network models, we can ask at least three distinct kinds of questions. The first type of question concerns structure and knowledge: to what extent can the organization of human semantic knowledge be explained in terms of general structural principles that characterize the connectivity of semantic networks? The second type concerns process and performance: to what extent can human performance in semantic processing tasks be explained in terms of general processes operating on semantic networks? A third type of question concerns the interactions of structure and process: to what extent do the processes of semantic retrieval and search exploit the general structural features of semantic networks, and to what extent do those structural features reflect general processes of semantic acquisition or development?

The earliest work on semantic networks attempted to confront these questions in an integrated fashion. Collins and Quillian (1969) suggested that concepts are represented as nodes in a tree-structured hierarchy, with connections determined by class-inclusion relations (Figure 1). Additional nodes for characteristic attributes or predicates are linked to the most general level of the hierarchy to which they apply. A tree-structured hierarchy provides a particularly economical system for representing default knowledge about categories, but it places strong constraints on the possible extensions of predicates – essentially, on the kinds of knowledge that are possible (Keil, 1979; Sommers, 1971). Collins and Quillian proposed algorithms for efficiently searching these inheritance hierarchies to retrieve or verify facts such as "robins have wings", and they showed that reaction times of human subjects often seemed to match the qualitative predictions of this model. However, notwithstanding the elegance of this picture, it has severe limitations as a general model of semantic structure. Inheritance hierarchies are clearly appropriate only for certain taxonomically organized concepts, such as classes of animals or other natural kinds. Even in those ideal cases, a strict inheritance structure seems not to apply except for the most typical members of the hierarchy (Carey, 1985; Collins and Quillian, 1969; Rips, Shoben, & Smith, 1973; Sloman, 1998).

Subsequent work on semantic networks put aside the search for general structural principles of knowledge organization and instead focused on elucidating the mechanisms of semantic processing in arbitrarily structured networks. The network models of Collins and Loftus (1975), for instance, are not characterized by any kind of large-scale structure such as a tree-like hierarchy. In terms of their large-scale patterns of connectivity, these models are essentially unstructured, with each word or concept corresponding to a node and links between any two nodes that are directly associated in some way (Figure 1B). Quantitative models of generic associative networks, often equipped with some kind of spreading-activation process, have been used to predict performance in a range of experimental memory retrieval tasks and to explain various priming and interference phenomena (Anderson, 2000; Deese, 1965; Collins & Loftus, 1975; Nelson, McKinney, Gee, & Janczura, 1998).

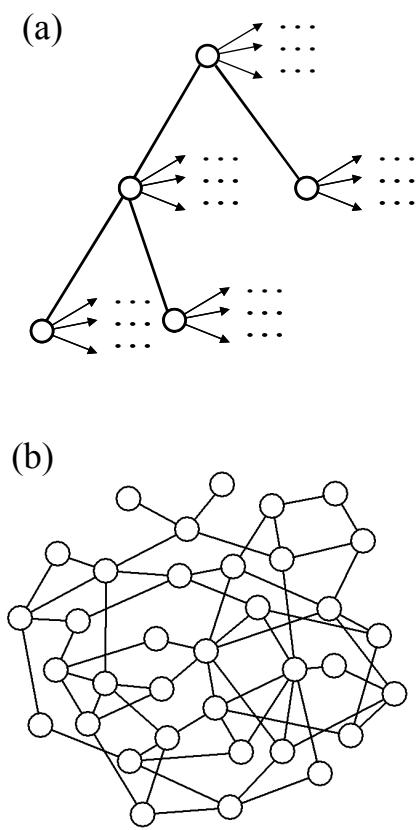

Figure 1. Proposed large-scale structures for semantic networks: (a), a tree-structured hierarchy (e.g., Collins & Quillian, 1969); (b), an arbitrary graph (e.g., Collins & Loftus, 1975).

As a result of research in this tradition, there is now a fair consensus about the general character of at least some of the processes involved in the formation and search of semantic memory (Anderson, 2000). By contrast, there is relatively less agreement about general principles governing the large-scale structure of semantic memory, or how that structure interacts with processes of memory search or knowledge acquisition. Typical textbook pictures of semantic memory still depict essentially arbitrary networks, such as Figure 1B, with no distinctive large-scale structures. The implications for semantic network theories of meaning are not good. Under the semantic net view, meaning is inseparable from structure: the meaning of a word or concept is in large part defined by the other words or concepts it connects to. Thus, without any general structural principles, the semantic net paradigm offers little or no general insights into the nature of semantics.

In this paper, we argue that there are in fact compelling general principles governing the structure of network representations for natural language semantics, and that these structural principles have potentially significant implications for the processes of semantic growth and memory search. We stress from the outset that these principles are far from composing a network theory of meaning. Our goal here is merely to study some of the general structural properties of semantic networks that may ultimately form part of the groundwork for such a theory.

The principles we propose are not based on any fixed structural motif such as the tree-structured hierarchy of Collins and Quillian (1969). Rather, they are based on statistical regularities that we have uncovered via graph-theoretic analyses of previously described semantic networks. We look at the distributions of several statistics calculated over nodes, pairs of nodes, or triples of nodes in a semantic network: the number of connections per word, the length of the shortest path between two words, and the percentage of a node's neighbors that are themselves neighbors. We show that semantic networks, like many other natural networks (Watts & Strogatz, 1998), possess a *small-world* structure characterized by the combination of highly clustered neighborhoods and a short average path-length. Moreover, this small-world structure seems to arise from a *scale-free* organization, also found in many other systems (Barabási & Albert, 1999; Strogatz, 2001), in which a relatively small number of well-connected nodes serve as hubs and the distribution of node connectivities follows a power function. Figure 7 shows an example of a small-world, scale-free network.

These statistical principles of semantic network structure are quite general in scope, in at least two senses. First, they apply on average to all words in the language, regardless of syntactic class or semantic domain. Second, they appear to hold for semantic network representations constructed in very different ways, whether from the word associations of naive subjects (Nelson, McEvoy & Schreiber, 1999) or the considered analyses of linguists (Roget, 1911; Miller, 1995). At the same time, these regularities do not hold for many popular models of semantic structure, including both hierarchical or randomly (arbitrarily) connected networks (Figures 1A and 1B), as well as high-dimensional vector space models such as Latent Semantic Analysis (Landauer & Dumais, 1997). These principles may thus suggest directions for new modeling approaches, or for extending or revising existing models. Ultimately, they may help to determine which classes of models most faithfully capture the structure of natural language semantics.

As in studies of scale-free or small-world structures in other physical, biological, or social networks (Albert, Jeong, & Barabási, 2000; Barabási & Albert, 1999; Watts & Strogatz, 1998), we will emphasize the implications of these distinctive structures for some of the crucial processes that operate on semantic networks. We suggest that these structures may be consequences of the developmental mechanisms by which connections between words or concepts are formed – either in language evolution, language acquisition, or both. In particular, we show how simple models of network growth can produce close quantitative fits to the statistics of real semantic networks, based only on plausible abstract principles with no free numerical parameters.

In our models, a network acquires new concepts over time and connects each new concept to a subset of the concepts within an existing neighborhood, with the probability of choosing a particular neighborhood proportional to its size. This growth process can be viewed as a kind of semantic differentiation, in which new concepts correspond to more specific variations on existing concepts and highly complex concepts (those with many connections) are more likely to be differentiated than simpler ones. It naturally yields scale-free small-world networks, such as the one shown in Figure 6.

Our models also make predictions about the time-course of semantic acquisition, because the order in which words are acquired is crucial in determining their connectivity. Words that enter the network early are expected to show higher connectivity. We verify this relationship experimentally with age-of-acquisition norms (Gilhooly & Logie, 1980; Morrison, Chappell, & Ellis, 1997) and explain how it could account for some puzzling behavioral effects of age of acquisition in lexical decision and naming tasks, under plausible assumptions about search mechanisms in semantic memory.

## Basic Concepts from Graph Theory

We begin by defining some terminology from graph theory and briefly introducing the statistical properties that we will use to describe the structure of semantic networks.[1] Underlying every semantic network is a *graph,* consisting of a set of *nodes* (also called *vertices*) and a set of *edges* or *arcs* that join pairs of nodes. The number of nodes in the network is denoted by *n*. An edge is an undirected link between two nodes and a graph containing only edges is said to be *undirected*. An arc is a directed link between two nodes and a graph containing only arcs is said to be *directed*. Every directed graph corresponds naturally to an undirected graph over the same nodes, obtained by replacing each arc with an edge between the same pair of nodes.

Two nodes that are connected by either an arc or edge are said to be *neighbors*; a *neighborhood* is a subset of nodes consisting of some node and all of its neighbors. When the network is directed, the *in-degree* and *out-degree* of a node refer to the numbers of arcs incoming to or outgoing from that node, respectively. The variables $k_i^{in}$ and $k_i^{out}$ denote the in- and out-degree of node *i* respectively. When the network is undirected, the in-degree and out-degree are always equal and we refer to either quantity as the *degree* of a node. We write the degree of node *i* as $k_i$. We will also use $k_i$ with directed networks to denote the degree of node *i* in the corresponding undirected network (i.e., the total numbers of neighbors of node *i*).

In an undirected graph, a *path* is a sequence of edges that connects one node to another. In a directed graph, a (directed) path is a set of arcs that can be followed along the direction of the arcs from one node to another. We can also talk about undirected paths in a directed graph, referring to the paths along edges in the corresponding undirected graph, but by default any reference to paths in a directed network will assume the paths are directed. For a particular path from node *x* to node *y*, the *path length* is the number of edges (in an undirected graph) or arcs (in a directed graph) along that path. We refer to the *distance* between *x* and *y* as the length of the shortest path connecting them.[2] In a *connected* graph, there exists an undirected path between any pair of nodes. A directed graph is said to be *strongly connected* if between any pair of nodes there exists a path along the arcs. A (strongly) *connected component* is a subset of nodes that is (strongly) connected.

We will characterize the structure of semantic networks primarily in terms of four statistical features defined using the above terminology. These quantities are the average distance *L*, the diameter *D*, the clustering coefficient *C*, and the degree distribution *P(k)*. *L* and *D* are closely related: *L* refers to the average of the shortest path lengths between all pairs nodes in a network, while *D* refers to the maximum of these distances over all pairs of nodes. In other words, at most *D* steps are needed to move from any node to any other, but on average only *L* steps are required.[3]

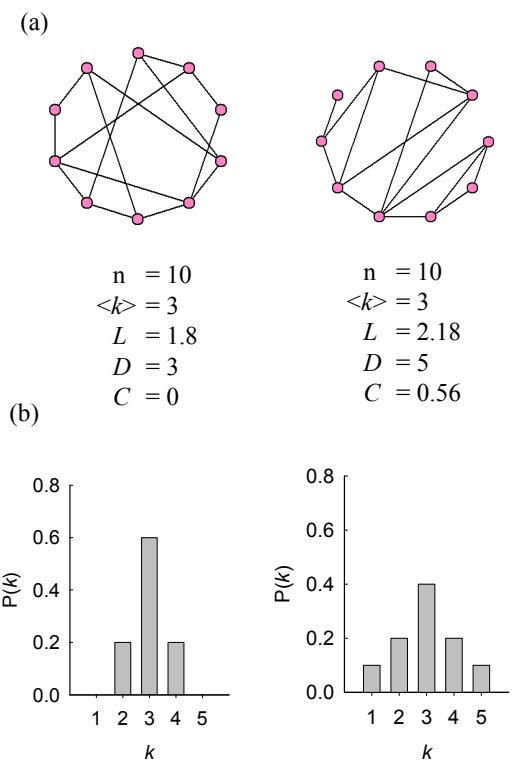

**Figure 2**. An illustration of the graph-theoretic properties that we will apply to semantic networks. (a) Two networks with equal numbers of nodes and edges. For both networks, the variables *n* (number of nodes) and <*k*> (average degree, i.e., average number of edges) are shown as well as the statistical properties *L* (average shortest path length), *D* (diameter) and *C* (clustering coefficient). Note that the two networks in have different clustering coefficients even though they have the same *n* and <*k*>. (b) The degree distributions corresponding to the two networks above. Both networks show the typical pattern for random graphs: approximately bell-shaped distributions with tails that decay exponentially as *k* increases.

The clustering coefficient *C* and the degree distribution *P(k)* are two different probabilistic measures of the graph connectivity structure. *C* represents the probability that two neighbors of a randomly chosen node will themselves be neighbors, or alternatively, the extent to which the neighborhoods of neighboring nodes overlap. Following Watts and Strogatz (1998), we calculate *C* by taking the average over all nodes *i* of the quantity

$$C_i = T_i \bigg/ \binom{k_i}{2} = 2T_i \big/ k_i(k_i-1) \qquad (1)$$

where $T_i$ denotes the number of connections between the neighbors of node *i*, and $k_i(k_i-1)/2$ is the number of connections that would be expected between *i*'s neighbors if they formed a fully connected subgraph. Because $T_i$ can never exceed $k_i(k_i-1)/2$, the clustering coefficient *C* is normalized to lie between 0 to 1, as required of a probability. When *C*=0, no nodes have neighbors that are also each others' neighbors. In a fully connected network (i.e., every node is connected to all other nodes), *C*=1. While the clustering coefficient is sensitive to the number of connections in a network, it is possible for two networks to have the same number of connections but different clustering coefficients (see Figure 2). Finally, note that because the definitions for $T_i$ and $k_i$ are independent of whether the connections are based on edges or arcs, the clustering coefficients for a directed network and the corresponding undirected network are equal.

The degree distribution *P(k)* represents the probability that a randomly chosen node will have degree *k* (i.e., will have *k* neighbors). For directed networks, we will concentrate on the distribution of in-degrees, although one can also look at the out-degree distribution. We estimate these distributions based on the relative frequencies of node degrees found throughout the network. The most straightforward feature of *P(k)* is the expected value <*k*> under *P(k)*. This quantity, estimated by simply averaging the degree of all nodes over the network, ranges between 0 and *n* (for a fully connected network of *n* nodes) and represents the mean density of connections in the network. More information can be gleaned by plotting the full distribution *P(k)* as a function of *k*, using either a bar histogram (for small networks, as in Figure 2), a binned scatterplot (for large networks, as in Figure 5), or a smooth curve (for theoretical models, as in Figure 3). As we explain in the next section, the shapes of these plots provide characteristic signatures for different kinds of network structure and different processes of network growth.

Figure 2 shows these statistical measures for two different networks with 10 nodes and 15 edges. These examples illustrate how networks equal in size and density of connections may differ significantly in their other structural features. Figure 2 also illustrates two general properties of random graphs – graphs which are generated by placing an edge between any two nodes with some constant probability $p$ independent of the existence of any other edge (a model introduced by Erdös and Réyni, 1960). First, for fixed $n$ and $<k>$, high values of C tend to imply high values of $L$ and $D$. Second, the degree distribution $P(k)$ is approximately bell-shaped, with an exponential tail for high values of $k$.[4] While these two properties hold reliably for random graphs, they do not hold for many important natural networks, including semantic networks in natural language. We next turn to a detailed discussion of the small-world and scale-free structures that do characterize natural semantic networks. Both of these structures can be thought of in terms of how they contrast with random graphs: small-world structures are essentially defined by the combination of high values of $C$ together with low values of $L$ and $D$, while scale-free structures are characterized by non-bell-shaped degree distributions, with power-law (rather than exponential) tails.

## Small-World and Scale-Free Network Structures

Interest in the small-world phenomenon originated with the classic experiments of Milgram (1967) on social networks. Milgram's results suggested that any two people in the United States were, on average, separated by only a small number of acquaintances or friends (popularly known as "six degrees of separation"). While the finding of very short distances between random pairs of nodes in a large sparsely connected network may seem surprising, it does not necessarily indicate any interesting structure. This phenomenon occurs even in the random graphs described above, where each pair of nodes is joined by an edge with probability $p$. When $p$ is sufficiently high, the whole network becomes connected and the average distance $L$ grows logarithmically with $n$, the size of the network (Erdös & Réyni, 1960).

Watts and Strogatz (1998) sparked renewed interest in the mathematical basis of the small-world phenomenon with their study of several naturally occurring networks: the power grid of the western United States, the collaboration network of international film actors and the neural network of the worm *C. Elegans*. They showed that while random graphs with comparable size $n$ and mean connectivity $<k>$ describe very well the short path-lengths found in these networks, they also exhibit clustering coefficients $C$ that are orders of magnitude smaller than those observed in the real networks. In other words, natural small-world networks somehow produce much shorter internode distances than would be expected in equally dense random graphs, given how likely it is that the neighbors of a node are also each other's neighbors (see note 4). For the remainder of this paper, we use the term *small-world structure* to refer to this combination of short average path-lengths $L$ and relatively high clustering coefficients $C$ (by comparison with equally dense random graphs).

Small-world structures have since been found in many other networks (reviewed in Strogatz, 2001), including the world-wide-web (WWW; Adamic, 1999; Albert, Jeong, & Barabási, 1999), networks of scientific collaborators (Newman, 2001), and metabolic networks in biology (Jeong, Tombor, Albert, Oltval & Barabási, 2000). Watts and Strogatz (1998) proposed a simple abstract model for the formation of small-world structures, in which a small number of the connections in a low-dimensional regular lattice are replaced with connections between random pairs of nodes. The local neighborhood structure of the lattice leads to high clustering while the long-range random connections lead to very short average path-lengths.

Amaral, Scala, Barthélémy, and Stanley (2000) distinguish between different classes of small-world networks by measuring the degree distribution $P(k)$. In one class of networks, such as *C. Elegans* and the U. S. power grid, the degree distribution decays exponentially. This behavior is well described by random graph models or variants of the Watts and Strogatz model. In other systems, such as the WWW or metabolic networks, the degree distribution follows a power law (Barabási & Albert, 1999),

$$P(k) \approx k^{-\gamma} \qquad (2)$$

for values of γ typically between 2 and 4. Figure 3 illustrates the difference between power-law and exponential degree distributions. Intuitively, a power-law distribution implies that a small but significant number of nodes are connected to a very large number of other nodes, while in an exponential distribution, such "hubs" are essentially nonexistent. Because networks with power-law degree distributions have no characteristic scale of node degree, but instead exhibit all scales of connectivity simultaneously, they are often referred to as *scale-free structures* (Barabási & Albert, 1999). Power-law and exponential distributions can be differentiated most easily by plotting them in log-log coordinates (as shown in Figure 3b). Only a power-law distribution follows a straight line in log-log coordinates, with slope given by the parameter γ.

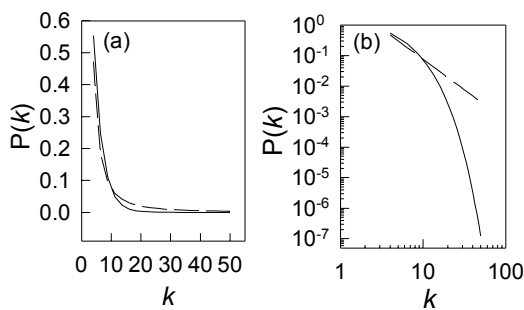

**Figure 3**. Two different kinds of degree distributions that may be observed in complex networks. (a) A power-law distribution (dashed) and an exponential distribution (solid), shown with the same linear scales as the histograms in Figure 2b. (b) When plotted with log-log scaling, the same distributions are more clearly differentiated, particularly in their tails.

Barabási and Albert (1999) have argued that the finding of power-law degree distributions places strong constraints on the process that generates a network's connectivity. They proposed an abstract model for scale-free network formation based on two principles that we explain in greater detail below: incremental growth and preferential attachment. This model yields power-law degree distributions, but it does not produce the strong neighborhood clustering characteristic of many small-world networks and the model of Watts and Strogatz (1998). In short, while the model of Watts and Strogatz (1998) naturally produces small-world structures and the model of Barabási & Albert (1999) naturally produces scale-free structures, neither of these approaches explains the emergence of both scale-free and small-world structures as have been observed in some important complex networks such as the WWW. There is currently a great deal of interest within the theoretical physics community in developing models of network formation that can capture both of these kinds of structures.

In the next section, we show that semantic networks, like the WWW, exhibit both small-world and scale-free structures. In the following section we introduce a model for network growth that is related to the Barabási and Albert model but that grows through a process of differentiation analogous to mechanisms of semantic development. This growth process allows our model to produce both small-world and scale-free structures naturally, with essentially no free parameters. The final section explores some of the psychological implications of this model and compares it to other frameworks for modeling semantic structure.

## Graph-Theoretic Analyses of Semantic Networks

We constructed networks based on three sources of semantic knowledge: free association norms (Nelson et al. 1999), WordNet (Fellbaum, 1998; Miller, 1995) and Roget's thesaurus (Roget, 1911). Although the processes that generated these data surely differ in important ways, we will see that the resulting semantic networks are similar in the statistics of their large-scale organization. To allow the application of conventional graph-theoretic analyses, we will construct these networks with all arcs and edges unlabeled and weighted equally. More subtle analyses that recognize qualitative or quantitative differences between connections would be an important subject of future work.

### Methods
Associative Network. A large free-association database involving more than 6000 participants was collected by Nelson et al. (1999). Over 5000 words served as cues (e.g. "cat") for which participants had to write down the first word that came to mind (e.g. "dog"). We created two networks based on these norms. In the directed network, two word nodes $x$ and $y$ were joined by an arc (from $x$ to $y$) if the cue $x$ evoked $y$ as an associative response for at least two of the participants in the database. In the undirected network, word nodes were joined by an edge if the words were associatively related regardless of associative direction. While the directed network is clearly a more natural representation of word associations, our other networks were both undirected, and most of the literature on small-world and scale-free networks has focused on undirected networks. Hence the undirected network of word associations provides an important benchmark for comparison. Figure 4a shows a small part of the undirected semantic network highlighting one of the shortest associative path of length 4 between VOLCANO and ACHE (there are many different shortest paths between these words). Figure 4b shows all shortest associative paths from VOLCANO to ACHE in the directed associative network.

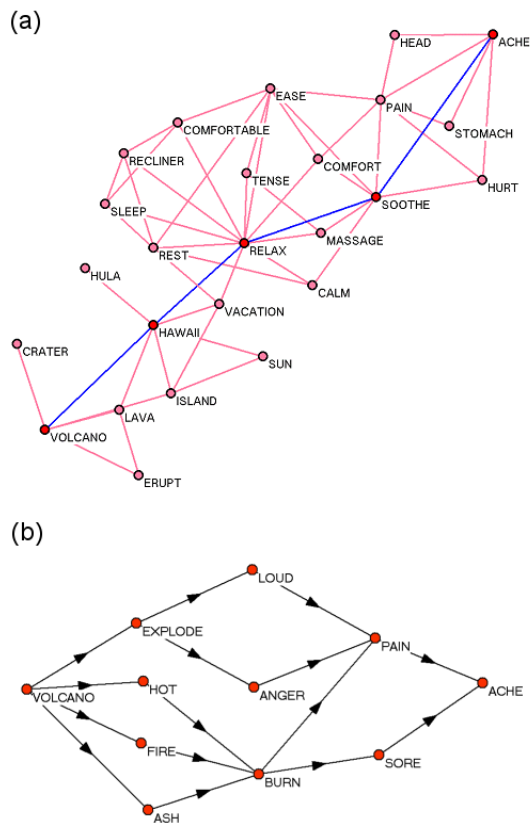

**Figure 4**. (a) Part of the semantic network formed by free association. The shortest path from VOLCANO to ACHE is highlighted. (b) all shortest directed paths from VOLCANO to ACHE.

Roget's Thesaurus (1911 edition). Based on the life long work of Dr. Peter Mark Roget (1779-1869), the 1911 edition includes over 29,000 words classified into 1000 semantic categories (ignoring several levels of subcategories). Roget's thesaurus can be viewed as a *bipartite graph*, a graph in which there are two different kind of nodes, word nodes and semantic category nodes, with connections allowed only between two nodes of different kinds. In this graph, a connection is made between a word and category node when the word falls into the semantic category.

WordNet. Somewhat analogous to Roget's thesaurus, but inspired by modern psycholinguistic theory, WordNet was developed by George Miller and colleagues (Fellbaum, 1998; Miller, 1995). The network contains 120,000+ word forms (single words and collocations) and 99,000+ word meanings. The basic links in the network are between word forms and word meanings. Multiple word forms are connected to a single word meaning node if the word forms are synonymous. A word form is connected to multiple word meaning nodes if it is polysemous.

Word forms can be connected to each other through a variety of relations such as antonymy (e.g., BLACK and WHITE). Word meaning nodes are connected by relations such as hypernymy (MAPLE is a TREE) and meronymy (BIRD has a BEAK). Although these relations such as hypernymy and meronymy are directed, they can be directed both ways depending on what relationship is stressed. For example, the connection between BIRD and BEAK can be from bird to beak because birds have beaks but also from beak to bird because a beak is part of a bird. Because there are no inherently preferred directions for these relationships, we will treat WordNet as an undirected graph.

**Results and analyses**

Our analysis of these semantic networks focuses on five properties: sparsity, connectedness, short path-lengths, high neighborhood clustering, and power-law degree distributions. The statistics related to these properties are shown in Table 1 (under the Data columns), and the estimated degree distributions for each network are plotted in Figure 5. To provide a benchmark for small-world analyses, we also computed the average shortest-path lengths ($L_{random}$) and clustering coefficients ($C_{random}$) for ensembles of random networks with sizes and connection densities equal to those observed in the three semantic networks. These random graphs were created by randomly rearranging connections in the corresponding semantic networks.[5]

Sparsity. For WordNet and Roget, the number of nodes can be separated into the number of word nodes and the number of class nodes (categories in Roget and word meanings in WordNet). For WordNet and Roget's Thesaurus, Table 1 lists $<k>$ (the average degree or average number of connections) separately for word and class nodes. Given the size of the networks and the number of connections, it can be observed that all three semantic networks are sparse: on average, a node is connected to only a very small percentage of other nodes. In the undirected associative network, a word is connected on average to only 22 (.44%) of the 5018 total number of words. The semantic networks of WordNet and Roget's thesaurus exhibit even sparser connectivity patterns.

Connectedness. Despite their sparsity, each of these semantic networks contains a single large connected component that includes the vast majority of nodes. In the directed associative network, the largest strongly connected component consists of 96% of all words (i.e., for this set of words, there is an associative path from any word to any other word when the direction of association is taken into account). In the undirected associative network, the whole network is connected. For both WordNet and Roget's thesaurus, the largest connected component consists of more than 99% of all words. We restricted all further analyses to these components.

Short Path-Lengths. All three networks display very short average path-lengths and diameters relative to the sizes of the networks[6]. For instance, in the undirected associative network, the average path-length ($L$) is only 3 while the maximum path length ($D$) is only 5. That is, at most 5 associative steps (independent of direction) separate any two words in the 5,000+ word lexicon. These short path lengths and small diameters are well-described by random graphs of equivalent size and density, consistent with Watts and Strogatz's (1998) findings for other small-world networks.

Neighborhood Clustering. The clustering coefficient $C$ for the associative network is well above zero, implying that the associates of a word tend also to be directly associated a significant fraction (approximately 1/6) of the time. The absolute value of $C$ appears much lower for WordNet, but that is primarily because the graph is much sparser. For both the associative network and WordNet, $C$ is several orders of magnitude larger than would be expected in a random graph of equivalent size and density ($C_{random}$). For Roget's thesaurus, the analysis of neighborhood clustering is more complex. Because the thesaurus is a bipartite graph, with connections between word and class nodes but not between nodes of the same type, the neighbors of a word node can never themselves be neighbors. In order to define a meaningful measure of semantic neighborhood clustering in the thesaurus network, we converted the bipartite graph to a simple graph on the word nodes by connecting words if they shared at least one class in common. The clustering coefficient $C$ was then computed on this word graph and compared with the mean clustering coefficient $C_{random}$ computed in an analogous manner for an ensemble of random bipartite graphs with the same size and density as the original thesaurus network. As in the other two semantic networks, $C$ was substantially higher for the thesaurus than for the comparable random graphs.

Power-Law Degree Distribution. Figure 5 plots the degree distributions for the word nodes of each network in log-log coordinates, together with the best-fitting power functions (which appear as straight lines under the log-log scaling). For the directed associative network, the in-degree distribution is shown. As in conventional histograms, these

**Table 1.** Summary statistics for semantic networks and model outputs.

| Variable[1] | Type | Undirected Associative Network | | | | | Directed Associative Network | | | Roget | WordNet |
|---|---|---|---|---|---|---|---|---|---|---|---|
| | | Data | Model A | | Model B[2] | | Data | Model B | | Data | Data |
| $n$ | words | 5,018 | 5,018 | | 5,018 | | 5,018 | 5,018 | | 29,381 | 122,005 |
| | classes | - | - | | - | | - | - | | 1,000 | 99,642 |
| $<k>$ | words | 22.0 | 22.0 | | 22.0 | | 12.7 | 13.0 | | 1.7 | 1.6 |
| | classes | - | - | | - | | - | - | | 49.6 | 4.0 |
| $L$ | | 3.04 | 3.00 | (.012) | 3.00 | (.009) | 4.27 | 4.28 | (.030) | 5.60 | 10.56 |
| $D$ | | 5 | 5.00 | (.000) | 5.00 | (.000) | 10 | 10.56 | (.917) | 10 | 27 |
| $C$ | | .186 | .174 | (.004) | .173 | (.005) | .186 | .157 | (.003) | .875 | .0265 |
| $\gamma$ | | 3.01 | 2.95 | (.054) | 2.97 | (.046) | 1.79 | 1.90 | (.021) | 3.19 | 3.11 |
| $L_{random}$ | | 3.03 | - | | - | | 4.26 | - | | 5.43 | 10.61 |
| $C_{random}$ | | 4.35E-03 | - | | - | | 4.35E-03 | - | | .613 | 1.29E-04 |

Note: Standard deviations of 50 simulations given between parentheses.

(1) The following notation was used: $n$ (the number of nodes), $<k>$ (the average number of connections), $L$ (the average shortest path length), $D$ (the diameter of the network), $C$ (clustering coefficient), $\gamma$ (power law exponent for the distribution of the number of edges in undirected networks and incoming connections in directed networks), $L_{random}$ (the average shortest path length with random graph of same size and density), and $C_{random}$ (the clustering coefficient for a random graph of same size and density)

(2) In these simulations, the directed networks from model B were converted to undirected networks.

distributions were estimated by grouping all values of *k* into bins of consecutive values and computing the mean value of *k* for each bin. The mean value of each bin corresponds to one point in Figure 5. The boundaries between bins were spaced logarithmically to ensure approximately equal numbers of observations per bin.

For the three undirected networks, power functions fit the degree distributions almost perfectly. The exponents γ of the best-fitting power laws (corresponding to the slopes of the lines in Figure 5) were quite similar in all three cases, varying between 3.01 and 3.19 (see Table 1). The high-connectivity words at the tail of the power-law distribution can be thought of as the "hubs" of the semantic network. In word association, these hubs typically correspond to important general categories, such as GOOD, BAD, FOOD, LOVE, WORK, MONEY, and HOUSE. In WordNet, they typically correspond to polysemous verbs such as BREAK, RUN, and MAKE.

For the directed associative network, the in-degree distribution shows a slight deviation from the power-law form and the best-fitting power law has an exponent γ somewhat lower than 2. The out-degree of words in the directed associative network (not shown in Figure 5) is not power-law distributed, but instead has a single peak near its mean and exponential tails, similar to a normal or $\xi^2$ distribution. We focus on the in-degree distribution as opposed to the out-degree distribution, because the out-degree of a node in the word associative network is strongly dependent on specific details of how the word association experiment was conducted: the number of subjects that gave associative responses to that cue and the number of different associates that each subject was asked to generate for that cue. We will discuss the differences between the in- and out-degree distributions in word association in more detail below, when we describe the growing network model that can explain these differences.

**Discussion**

All of the semantic networks studied shared the distinctive statistical features of both small-world and scale-free structures: a high degree of sparsity, a single connected component containing the vast majority of nodes, very short average distances between nodes, high local clustering, and a power-law degree distribution (with exponent near 3 for the undirected networks). The fact that these properties held for all networks despite their origins in very different kinds of data demands some explanation. It is unlikely that the commonalities are simply artifacts of our analysis, because they are not found in random graphs or even in many complex networks from other scientific domains that have been subjected to the same kinds of analyses (Amaral et al., 2000; Watts & Strogatz, 1998). It is more reasonable to suppose that they reflect, at least in part, some abstract features of semantic organization. This structure must be sufficiently deep and pervasive to be accessed by processes as diverse as rapid free association by naive subjects (Nelson et al., 1999) and considered analysis by linguistic experts (Miller, 1995; Roget, 1911).

The similarities in network structure may also depend in part on the coarse grain of our analysis, which treats all words and all connections between words equally. Surely these simplifications make our analysis insensitive to a great deal of interesting linguistic structure, but they may also enable us to see the forest for the trees – to pick up on general properties of meaning in language that might be missed in more fine-grained but small-scale studies of particular semantic domains. A promising direction for future work would be to refine our analyses based on some minimal linguistic constraints. For instance, words or connections could first be segregated into broad syntactic or semantic classes, and then the same statistical analyses could be applied to each class separately. Many authors have suggested that there are systematic differences in the semantics of nouns and verbs (Gentner, 1981) or nouns and adjectives (Gasser & Smith, 1998), or different kinds verbs (Levine, 1993), and it would be interesting to see if those differences correspond to different large-scale statistical patterns. It would also be interesting to apply the same analyses to the semantic networks of different languages. We expect that the general principles of small-world and scale-free structures would be universal, but perhaps we would find quantitative variations in the clustering coefficients or power-law exponents resulting from different language histories.

Power laws in human language were made famous by Zipf (1965), but were in fact discussed as early as the 1930's by Skinner and probably others. After conducting the above analyses, we discovered some intriguing parallels with those classic studies. Zipf's best-known finding concerns the distribution of word frequencies, but he also found a power-law distribution for the number of word meanings (as given by the Thorndike-Century dictionary). That is, most words have relatively few distinct meanings, but a small number of words have many meanings. If we assume that a word's degree of connectivity is proportional to the number of its distinct meanings, then Zipf's "law of meaning" is highly consistent with our results here, including a power-law exponent of approximately 3 that best characterizes his distribution.[7] In our analysis of the in-degree distribution for the directed associative network, the best-fitting power-law exponent was somewhat lower than 2. This result was anticipated by Skinner (1937), who measured the distribution of the number of different associative responses to a much smaller set of cues than did Nelson et al. (1999). His plots show power-law distributions with a slope somewhat lower than 2, which is quite consistent with our findings for the in-degree distribution of the word association network.

Given the limited significance attributed to the work of Zipf and Skinner in most contemporary accounts of linguistic structure and processing, it is quite reasonable to ask whether the statistical regularities we have uncovered will prove to be any more important. Skinner's work on the associative basis of language has been discounted primarily on the grounds that it looked only at the surface forms of language, ignoring the unobservable representational structures that cause those observable forms to behave as they do (Chomsky, 1957, 1959). Our analysis, in contrast, examines both simple associations between surface forms (Nelson et al., 1999) and more complex relations between words mediated by unobservable classes (WordNet), and finds similar patterns in them. The unified theory that Zipf (1965) proposed to account for his power-law findings, based on principles of least effort, has fared somewhat better than Skinner's theories of language.

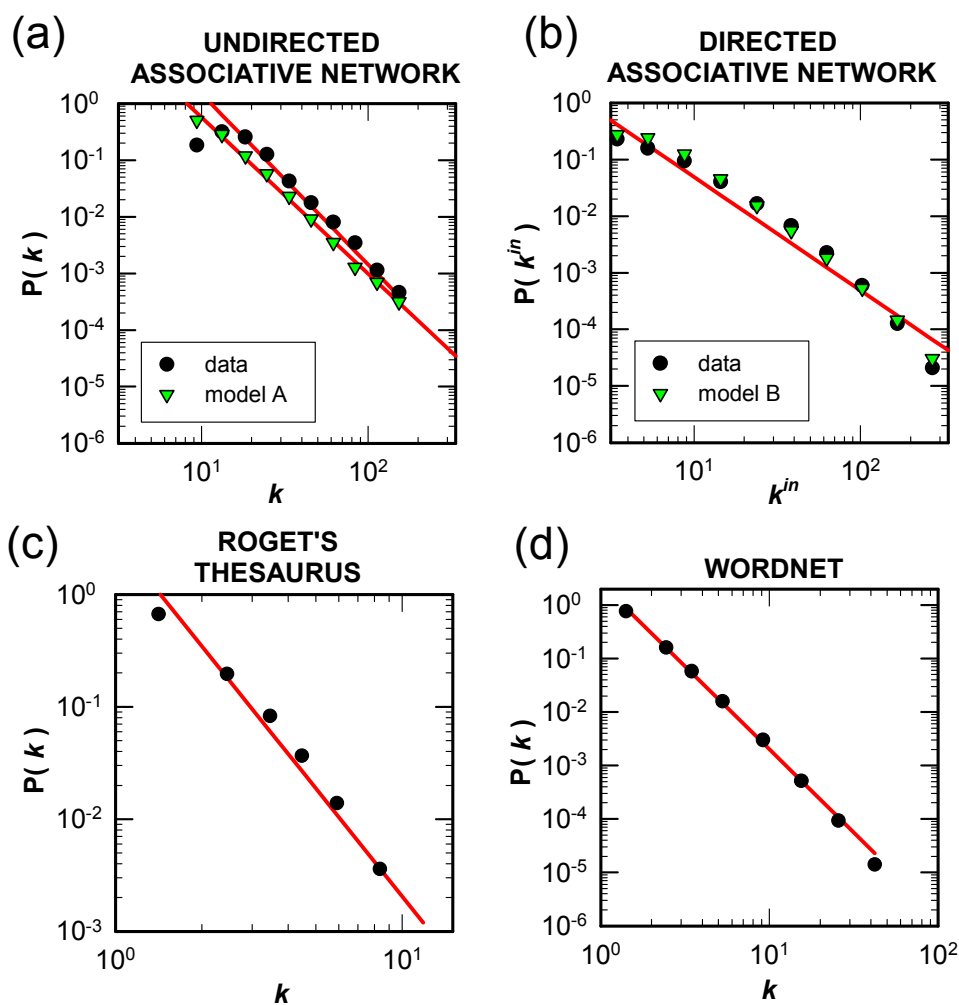

**Figure 5**. The degree distributions in three semantic networks, word association (a) and (b), Roget's thesaurus (c), and WordNet (d). All distributions are shown in log-log coordinates with the line showing the best fitting power law distribution. In panel (a), the undirected associative network is shown with the fit of model A: the undirected growing network model. In panel (b), the in-degree distribution of the directed associative network is shown along with the fit of model B: the directed growing network model. For Roget's thesaurus and WordNet, the degree distributions shown are for the word nodes only.

Yet it does not play a large role in contemporary computational studies of language, perhaps because its formulation is vague in many places and because many simple mathematical models have been subsequently proposed that can reproduce Zipf's most famous result – the power-law distribution of word frequencies – without offering any deep insights into how language works (Manning & Schutze, 1999). In contrast, the statistical properties we have identified are not predicted by the currently most popular accounts of semantic structure, nor by many other mathematical models of network formation. We have also attempted to develop mathematically precise and psychologically motivated models for the formation of semantic networks that are consistent with all of these constraints. These models are the subject of the following section.

## Growing Network Model

It has been argued by a number of researchers (Barabási and Albert, 1999; Kleinberg, Kumar, Raghavan, Rajagopalan, & Tomkins, 1999; Simon, 1955) that power-law distributions are a consequence of the characteristic ways that systems grow or develop over time. In particular, power laws in network degree distributions have been taken as the signature of a particular kind of network growth process known as *preferential attachment* (Barabási and Albert, 1999; see also Simon, 1955): nodes are added to the network successively, by connecting them to a small sample of existing nodes selected with probabilities proportional to their degrees. In other words, the more highly connected a node is, the more likely it is to acquire new connections.

Barabási and Albert (1999) proposed a simple model that directly instantiates the principle of preferential attachment: each target for a new node's connections is sampled independently from the set of all existing nodes, with probability proportional to its current degree. This model produces power-law degree distributions with an exponent of 3, much like those we observed for semantic networks. However, it does not yield clustering coefficients that are nearly as high as those as we observed. For instance, when the network size and density are comparable to the word association network, the model of Barabási and Albert (1999) yields values of $C$ around 0.02, much lower than the observed value of 0.186. Asymptotically, as network size grows to infinity, $C$ approaches 0 for this model, making it inappropriate for modeling small-world structures. Conversely, the classic model of small-world network formation, due to Watts and Strogatz (1998), does not capture the scale-free structures we have observed. In this section, we present an alternative model of preferential attachment that draws its inspiration from mechanisms of semantic development and that naturally produces both scale-free structures – with appropriate power-law exponents – and small-world structures – with appropriate clustering coefficients.

We should stress that in modeling the growth of semantic networks, our aim is not to describe in detail any specific psychological mechanism. Rather, we seek to capture at an abstract level the relations between the statistics reported in the previous section and the dynamics of how semantic structures might grow. Our model's most natural domain of applicability is to semantic growth within an individual – the process of lexical development – but it may also be applicable to the growth of semantic structures shared across different speakers of a language or even different generations of speakers – the process of language evolution.

We frame our model abstractly in terms of nodes – which may be thought of as words or concepts – and connections between nodes – which may be thought of as semantic associations or relations. Over time, new nodes are added to the network and probabilistically attached to existing nodes on the basis of three principles. First, following the suggestions of many researchers in language and conceptual development (Brown, 1958a,b; Carey, 1978, 1985; Clark, 1993, 2001; Macnamara, 1982; Slobin, 1973), we will assume that semantic structures grow primarily through a process of differentiation: the meaning of a new word or concept typically consists of some kind of variation on the meaning of an existing word or concept. Specifically, we assume that when a new node is added to the network, it differentiates an existing node by acquiring a pattern of connections that corresponds to a subset of the existing node's connections. Second, we assume that the probability of differentiating a particular node at each timestep is proportional to its current complexity – how many connections it has. Finally, we allow nodes to vary in a "utility" variable, which modulates the probability that they will be the targets of new connections. Utility variation is not necessary to produce any of the statistical features described in the previous section; it merely allows us to explore interactions between those features and aspects of word utility, such as usage frequency.

By focusing on the process of semantic differentiation, we do not mean to preclude a role for other growth processes. We have chosen to base our model on this single process strictly in the interests of parsimony. Incorporating additional processes would surely make the model more realistic but would also entail adding in more free parameters, corresponding to the relative weights of those mechanisms. Given that the data we wish to account for consists of only the few summary statistics in Table 1 and the degree distributions in Figure 5, it is essential to keep the number of free parameters to an absolute minimum. Our model for undirected networks (model A) has no free numerical parameters, while our model for directed networks (model B) has just one free parameter.

## Model A: the Undirected Growing Network Model

Let $n$ be the size of the network that we wish to grow, and $n(t)$ denote the number of nodes at time $t$. Following Barabási and Albert (1999), we start with a small fully connected network of $M$ nodes ($M << n$). At each time step, a new node with $M$ links is added to the network by randomly choosing some existing node $i$ for differentiation, and then connecting the new node to $M$ randomly chosen nodes in the semantic neighborhood of node $i$. (Recall that the neighborhood $H_i$ of node $i$ consists of $i$ and all the nodes connected to it.) Under this growth process, every neighborhood always contains at least $M$ nodes; thus a new node always attaches to the network by connecting to a subset of the neighborhood of one existing node. In this sense, the new node can be thought of as differentiating the existing node, by acquiring a similar but slightly more specific pattern of connectivity.

To complete the model, we must specify two probability distributions. First, we take the probability $P_i(t)$ of choosing node $i$ to be differentiated at time $t$ to be proportional to the complexity of the corresponding word/concept, as measured by its number of connections:

$$P_i(t) = \frac{k_i(t)}{\sum_{l=1}^{n(t)} k_l(t)} \quad (3)$$

where $k_i(t)$ is the degree (number of connections) of node $i$ at time $t$. The indices in the denominator range over all existing $n(t)$ nodes in the network. Second, given that node $i$ has been selected for differentiation, we take the probability $P_{ij}(t)$ of connecting to a particular node $j$ in the neighborhood of node $i$ to be proportional to the utility of the corresponding word/concept:

$$P_{ij}(t) = \frac{u_j}{\sum_{l \in H_i} u_l} \quad (4)$$

where the indices in the denominator range over all nodes in the neighborhood $H_i$. To explore the interaction between word frequencies and network structure, we may equate a word's utility with its usage frequency (e.g., Kucera & Francis, 1967). For simplicity, we may also take all utilities to be equal, in which case the connection probabilities are simply distributed uniformly over the neighborhood of node $i$:

$$P_{ij}(t) = \frac{1}{k_i(t)} \quad (5)$$

For each new node added to the network, we sample repeatedly from the distribution in (4) or (5) until $M$ unique nodes within the neighborhood of $i$ have been chosen. The new node is then connected to those $M$ chosen nodes. We continue adding nodes to the network until the desired network size $n$ is reached. The growth process of the model and a small resulting network with $n=150$ and $M=2$ is illustrated in Figure 6.

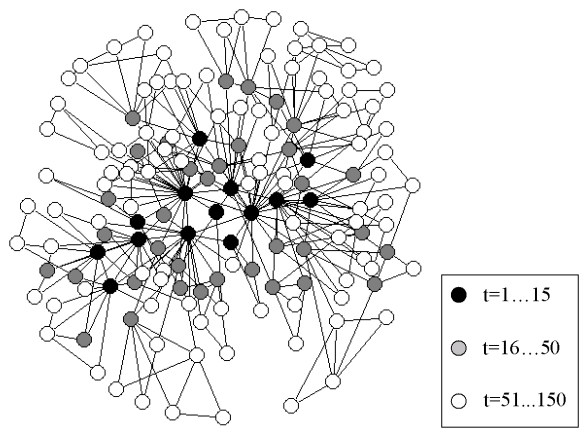

**Figure 6**. Illustration of the undirected growing network model with $n=150$ and $M=2$. The color of the nodes indicates the timestep at which the nodes were first inserted.

In our applications, $M$ and $n$ are not free to vary but are determined uniquely by the goal of producing a synthetic network comparable in size and mean density of connections to some real target network that we seek to model. The size $n$ is simply set equal to the size of the target network. The parameter $M$ is set equal to one-half of the target network's mean connectivity $<k>$, based on the following rationale. Each new node in the synthetic network is linked to *M* other nodes, and the network starts with a small fully connected subgraph of *M* nodes. Hence the average number of connections per node in the synthetic network is <*k*> = 2*M* + *M* (*M*-1)/*n*, which approaches 2*M* as *n* becomes large.

## Model B: the Directed Growing Network Model

Our model for growing directed networks is practically identical to the model for undirected networks, with the addition of a mechanism for orienting the connections. The initial fully connected network with *M* nodes now contains an arc from each node to every other node. The probability of choosing node *i* to be differentiated at a particular time *t* is still given by Equation (3), with the degree of a node in the directed graph defined as the sum of its in- and out-degrees, $k_i = k_i^{in} + k_i^{out}$. Each new node still makes connections to *M* existing nodes within the neighborhood of the node it differentiates, and those nodes are still sampled randomly according to Equations (4) or (5). The main novelty of Model B is that now each new connection may point in one of two possible directions, towards the existing node or towards the new node. We make the simplifying assumption that the direction of each arc is chosen randomly and independently of the other arcs, pointing towards the older node with probability $\alpha$ and towards the new node with probability (1-$\alpha$). The value of $\alpha$ is a free parameter of the model. Whether connections represent superficial associations or deep semantic dependencies, it seems plausible that they should typically point from the new node towards the previously existing node, rather than the other way around. Hence we expect the best fitting value of $\alpha$ to be significantly greater 0.5, and perhaps just slightly less than 1.0.

**Results and Analyses**

In principle, we could compare the products of these models with any of the real semantic networks analyzed above. However, the computational complexity of the simulation increases dramatically as the network grows and it has not been practical for us to conduct systematic comparisons of the models with networks as large as WordNet or Roget's thesaurus. We have carried out a thorough comparison of Models A and B with the undirected and directed word association networks, respectively, and we report those results below. In all of these simulations, we set *n*=5018 to match the number of words in the free-association database. We set *M*=11 in Model A and *M*=13 in Model B to ensure that the resulting synthetic networks would end up with approximately the same density as the corresponding word association networks. We explored two different settings for the node utilities, one with all utilities equal and the other with utilities determined by word frequency, according to $u_i = \log(f_i + 1)$. The word frequencies $f_i$ were estimated from the Kucera & Francis (1967) counts of words in a large sample of written text. Because *M* and *n* were set to match the size and density of the associative networks and the utilities were set uniformly or by observed word frequencies, there were no free numerical parameters in these mechanisms. The only free parameter occurred in the directed model: we varied $\alpha$ and obtained best results near $\alpha$ = 0.95, corresponding to the reasonable assumption that on average, 19 out of 20 new directed connections point from a new node towards an existing node.

We evaluated the models by calculating the same statistical properties (Table 1) and degree distributions (Figure 5) discussed above for the real semantic networks. Because the growing models are stochastic, results vary from simulation to simulation. All the results we describe are averages over 50 simulations. These results are juxtaposed with the analogous data from the corresponding real semantic networks, in both Table 1 and Figure 5.

For all of the statistics summarized in Table 1, both models achieved close fits to the corresponding real semantic networks. The mean statistics from 50 model runs were always within 10% of the corresponding observed values, and usually substantially closer. Figure 5 shows that the degree distributions of the models matched those of the real semantic networks, both qualitatively and quantitatively. Model A, corresponding to the undirected associative network, produced a degree distribution with a perfect power-law tail and exponent near 3. Model B, corresponding to the directed associative network, produced an approximately power-law distribution, with a slight downward inflection and an exponent somewhat less than 2. All of these results were very similar regardless of whether utilities were taken to be equal (as shown in Table 1 and Figure 5) or variable according to the Kucera-Francis frequency distribution.

We also checked that the directed network model would reproduce the results of the undirected model when all directed links were converted to undirected links, which corresponds more accurately to the process by which the real undirected semantic network was constructed. We simulated model B with *M*=11 and *a*=0.95 and converted all arcs to edges at the end of each simulation. The results (shown in Table 1) were almost identical to those produced by model A. This is to be expected, because of the particularly simple mechanism for choosing the directionality of new connections in model B. The choice of directionality is made independently for each link and does not influence the other processes in the model. Removing the edge directionalities from Model B at the output stage thus renders it mechanistically equivalent to model A.

Model B is similar to a class of models for growing directed networks described by Kumar and colleagues (Kumar et al., 2000b; see also, Kleinberg et al., 1999; Kumar et al., 2000a). In the approach of Kumar et al., each new node is connected to the network with a constant number of outgoing arcs. The destination of each new arc is chosen by a two-step process: with probability $\alpha$, the destination is chosen at random from all nodes in the network; with probability 1-$\alpha$, the new arc copies the destination of one outgoing arc from a randomly selected existing node. Like our model B, this process also produces power-law degree distributions and high clustering coefficients.

## Discussion

The qualitative features of the simulated networks make sense in light of the differentiation mechanism we have hypothesized for semantic network growth. Short average path-lengths occur because of the presence of hubs in the network; many shortest paths between arbitrary nodes $x$ and $y$ involve one step from $x$ to a hub and from the hub to node $y$. These hubs, and more generally, the power-law degree distributions, result from a version of the preferential attachment principle. At any given timestep, nodes with more connections are more likely to receive new connections through the process of semantic differentiation, because they belong to more semantic neighborhoods, and those neighborhoods on average will be more complex (making them more likely to be chosen for differentiation, by Equation (3). The models produce high clustering coefficients because new connections are made only into existing semantic neighborhoods. This ensures high overlap in the neighborhoods of neighboring nodes.

It is not so easy to explain a priori the close quantitative match between the statistics and degree distributions of the word association networks and those of our simulated networks with comparable size and connection density. The fact that these results were obtained with no free parameters, in model A, or one free parameter set to a reasonable value, in model B, gives us reason to believe that something like the growth mechanisms instantiated in the models may also be responsible for producing the small-world and scale-free structures observed in natural-language semantic networks.

Our models are, at best, highly simplified abstractions of the real processes of semantic growth. This simplicity is necessary given how little we know about the large-scale structures of real semantic networks – essentially, no more than is summarized in Table 1 and Figure 5. By pairing down the details of semantic growth to a simple mechanism based on differentiating existing words or concepts, we have been able to provide a unifying explanation for several nontrivial statistical structures.

There are clearly many ways in which our models could be made more realistic. Currently, new nodes always form their connections within a single semantic neighborhood, and new connections are added only when new nodes are added – never between two existing nodes. It would not be hard to remove these constraints, but it would necessitate additional free parameters governing the probability of making connections outside of a neighborhood (or in two or more neighborhoods) and the probability of adding a new connection between existing nodes. Removing these constraints would also make the models more flexible in terms of the kind of data they can fit; currently, the clustering coefficient and the shape and slope of the degree distribution are uniquely determined by the size and density of the network. It would also be possible to build models with different kinds of nodes and different kinds of connections, perhaps governed by different principles of growth. This complexity could be particularly important for modeling the network structure of WordNet or Roget's thesaurus, which are based on a distinction between word nodes and class nodes. A thorough and rigorous explorations of the many possible models for semantic network growth should probably be deferred until we acquire more and better data about different kinds of semantic structures, and the computational resources to model the larger networks that we have already analyzed.

Finally, we acknowledge that we have been deliberately ambiguous about whether our model of semantic growth is meant to correspond to the process of language development within an individual's life span, or the process of language evolution across generations of speakers, or both. Although the mechanism of our model was primarily inspired by the language development literature (Brown, 1958a,b; Carey, 1978, 1985; Clark, 1993, 2001; Macnamara, 1982; Slobin, 1973), we think that some kind of semantic differentiation process is also a plausible candidate for how new word meanings are formed between individuals. Clearly these two processes are coupled, as the critical period of language acquisition is a principal locus of cross-generational linguistic change (Pinker, 1994). In future work, we hope to relate our modeling efforts more closely to the knowledge base of historical linguistics, as well as recent mathematical models of language evolution (Niyogi & Berwick, 1997; Nowak & Krakauer 1999).

## Psychological Implications of Semantic Growth

Our proposal that the large-scale structure of semantic networks arises from the mechanisms of semantic growth carries more general implications for psycholinguistic theories beyond merely accounting for the graph-theoretic properties described above. In this final section, we focus on two issues: the viability of non-growing semantic representations and the causal relationships between a network's growth history, semantic complexity, and memory search behavior.

## Power-law distributions and semantic growth

Conventional static views of semantic network organization – as either a hierarchical tree or an arbitrary graph – are consistent with the existence of short paths between any two nodes, but they do not predict either the small-world neighborhood clustering or the scale-free degree distributions that are found in real semantic networks. We have interpreted this constellation of features as the signature of a particular kind of network growth process, but we have not ruled out the possibility that some other kind of static semantic representation –
perhaps not based on a network model at all – may be give rise to these structures. A comprehensive survey of all previous semantic models is beyond the scope of this article, but we will explore one currently popular alternative representation based on the analysis of co-occurrences of words in a large corpus of text.

Latent Semantic Analysis (LSA; e.g., Landauer & Dumais, 1997; Landauer, Foltz, & Laham, 1998) has been proposed as a general theory for the representation and processing of semantic information. By analyzing the co-occurrence statistics of words across a large number of contexts in a corpus (where context is defined as a set of a few hundred words about a specific topic), the meanings of words can be represented by vectors in a high-dimensional linear vector space. The semantic similarity between words can then be determined by some measure of affinity in that space, such as the Euclidean distance, the inner product or the cosine of the angle between two vectors. Landauer and Dumais (1997) have shown that the local neighborhoods in semantic space successfully capture some subtle semantic relations. The question here is whether LSA captures other important structural features that we observe in semantic networks, such as the presence of hubs or a power-law distribution of neighborhood sizes.

To compare LSA representations with our semantic networks, we need some way of discretizing the continuous LSA space so that we can talk about the set of neighbors of a word rather than just the distances between words. In keeping with the spirit of previous LSA analyses (Landauer & Dumais, 1997;

**Table 2**. Statistical properties of networks constructed by LSA semantic spaces for different word sets and different dimensionalities.

| | Words Association Words | | | Most Frequent Words | | | All Words | | |
|---|---|---|---|---|---|---|---|---|---|
| | d | | | d | | | d | | |
| Variable | 50 | 200 | 400 | 50 | 200 | 400 | 50 | 200 | 400 |
| $n$ | 4,956 | 4,956 | 4,956 | 4,956 | 4,956 | 4,956 | 92,408 | 92,408 | 92,408 |
| $\tau$ | .614 | .338 | .225 | .608 | .332 | .221 | .614 | .338 | .225 |
| $<k>$ | 22.3 | 22.3 | 22.3 | 22.3 | 22.3 | 22.3 | 219.4 | 201.1 | 209.7 |
| $L$ | 4.83 | 4.02 | 3.70 | 4.77 | 3.86 | 3.62 | - | - | - |
| $D$ | 12 | 9 | 8 | 11 | 8 | 7 | - | - | - |
| $C$ | .456 | .391 | .298 | .454 | .354 | .274 | - | - | - |
| $\gamma$ | 1.07 | 1.08 | 1.03 | 1.03 | .64 | .34 | .76 | .40 | .06 |

*Note: empty cells in this table correspond to variables that could not be computed due to computational constraints; d = dimensionality of the vector space; $\tau$ = similarity threshold on the cosine of the angle for connecting two words*

Landauer, Foltz, & Laham, 1998), we create local neighborhoods by thresholding a continuous measure of dissimilarity defined by the angle between two vectors: two words are treated as neighbors if the cosine of the angle between their vectors exceeds some threshold $\tau$. By varying the subset of words included in the analysis and the threshold $\tau$, the size $n$ and mean connectivity $<k>$ of the LSA network can be made to correspond with those of the other networks we have studied.

We examined the LSA vector representation for three different subsets of the TASA corpus (Landauer, Foltz, & Laham, 1998): (a) all words ($n$=4956) which also occurred in the word association database; (b) the 4956 most frequently occurring words in the TASA corpus; and (c) all 92,408 words in the standard LSA database. The first set of words allowed us to compare the LSA network directly with the word associative network and the outputs of our growing network model. The second set of words provided an important control for the first set, to see if the results are dependent on the frequencies of the words as well as the particular set of words. The third set allowed us to see if results with the smaller vocabularies scale up to LSA networks comparable in size to WordNet or Roget's thesaurus. We also varied the dimensionality, $d$, of the LSA vector representation.

For word sets (a) and (b), we could carry out all the same analyses on the LSA networks formed by the thresholding procedure that we previously performed with the word associative network and the outputs of our model. We used thresholds $\tau$ that led to approximately the same $<k>$ as in the undirected word association network and our model A. For word set (c), computational constraints prevented us from calculating the statistics $L$, $D$, and $C$, but we were able to estimate their degree distributions. For consistency, we used thresholds $\tau$ identical to the thresholds used to construct the LSA network based on word set (a).

For all word sets, we used thresholds $\tau$ that led to approximately the same $<k>$ as in the undirected word association network and our model A. For word sets (a) and (b), we could carry out all the same analyses on the LSA networks formed by this thresholding procedure that we previously performed with the word associative network and the outputs of our model. For word set (c), computational constraints prevented us from calculating the statistics $L$, $D$, and $C$, but we were able to estimate their degree distributions.

Table 2 presents the statistics for these networks obtained for three different dimensionalities, $d = 50$, 200, and 400. (Typical psychological applications of LSA (Landauer & Dumais, 1997) set $d$ at approximately 300.) For both word-sets (a) and (b) and all three dimensionalities, the networks showed somewhat higher path lengths and clustering coefficients than observed in the associative networks or our model.[8] However, the LSA network statistics are qualitatively comparable and could be reasonably characterized as a small-world structure, with much higher clustering and shorter path-lengths than would be expected in a random network of equal size and density.

The LSA networks look much less plausible when the degree distributions are plotted. For word set (a), power-law scaling does not emerge at any dimensionality (Figure 7a). At high dimensionalities, the far tails of these distributions do look vaguely power-like, but it is difficult to confirm a strong linear trend for so few points, and the distributions clearly differ from those in Figure 5, where the power-law scaling holds over the vast majority of the data. We constructed LSA networks for word sets (b) and (c) in order to check whether a scale-free structure would emerge more clearly if we focused on just the most frequent words in the language, or if we included as many words as possible. However, we found even less evidence for a power-law degree distribution in these cases (Figures 7b and 6c). Because the distributions are curved in log-log coordinates, it becomes difficult to interpret the slope of the best fitting line ($\gamma$) in these plots. In Table 1, we nevertheless show the best-fitting values of $\gamma$ to highlight this key difference between the LSA models and the real semantic networks.

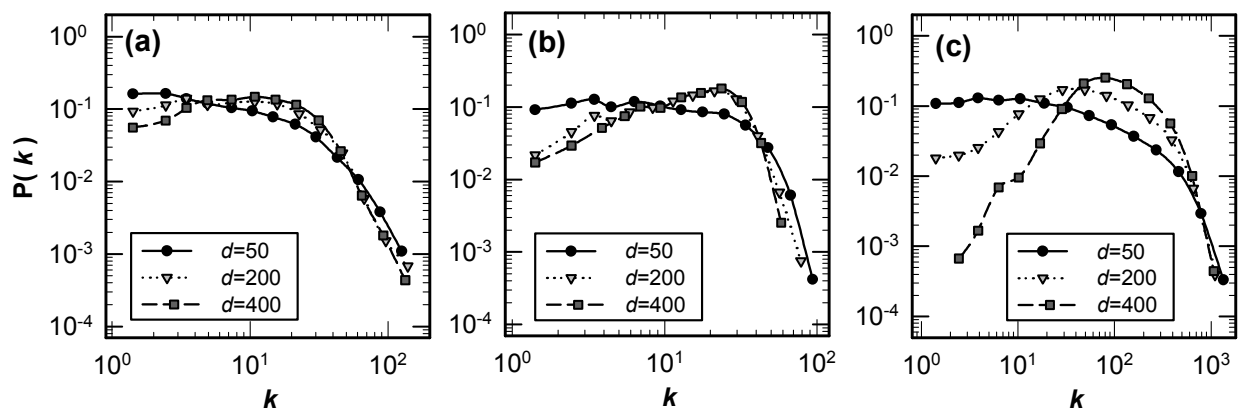

**Figure 7**. The degree distributions for networks based on thresholded LSA spaces for different dimensionalities. Panels (a), (b), and (c) correspond to simulations based on different subsets of words in the TASA corpus: the 4956 words found in the word association database (Nelson et al., 1999), the 4956 most frequent words (according to Kucera & Francis, 1967), and the complete set of 92,408 words, respectively.

We draw two tentative conclusions from our failure to find power-law scaling in the size distributions of semantic neighborhoods in LSA. First, the scale-free connectivity structures we have observed in a variety of real semantic networks are indeed nontrivial features, placing strong constraints on the mechanisms of network formation. Second, LSA and related co-occurrence-based models of meaning may need to be revised in some way. The appropriate revision may be an extension, such as adding some kind of dynamic growth process based on differentiation, or a more radical step, such as replacing the relatively unstructured, isotropic representational substrate of a Euclidean vector space with some more structured framework, such as a network.

There are several reasons to think that the constraint of scale-free connectivity will pose a challenge for many vector-space models of semantics, regardless of whether they are constructed through the machinery of LSA or some other procedure. Just as arbitrary graphs do not produce a power-law distribution of neighborhood sizes, neither do generic configurations of points in Euclidean space. For instance, if a large number of points are randomly distributed in a hypercube and pairs of points are connected as neighbors whenever they are within some small Euclidean distance $\varepsilon$ of each other, the number of neighbors per point will follow a Poisson distribution (Stoyan, Kendall, & Mecke, 1987). This distribution is just a limiting case of the binomial distribution that describes neighborhood sizes in a random graph; both distributions have exponential tails that show up as nonlinear in log-log coordinates. In a similar vein, Tversky and Hutchinson (1986) identified certain geometric properties of Euclidean-space semantic representations that are not consistent with human similarity judgments. In particular, they argued that Euclidean geometry – particularly in low dimensions – places strong constraints on the maximum number of nearest neighbors that any point may have, and that these constraints are not satisfied in conceptual domains with even very elementary forms of non-Euclidean structure, such as a taxonomic hierarchy.

It remains to be seen whether any static model can predict, in a nontrivial way, all of the large-scale structures that we observe in semantic networks and our growing network models. It may be relatively easy to hand-design a static representation with these properties, without answering the real causal questions. For example, consider a representation based on a large number of binary features (with each feature either present or absent for each word). Suppose that the features are chosen specifically so that the number of features per word is power-law distributed, and that two words are considered neighbors if they share at least one feature. Then the number of neighbors per word may also show a scale-free distribution, simply because the more features a word has, the more neighbors it will have. However, such an account only begs the question of where the power-law distribution of features comes from in the first place. Our growing network model, in contrast, provides a principled explanation for the origin of both small-world and scale-free structures, in the process of progressive semantic differentiation.

## Age of acquisition, Word Frequency, and Centrality

The core assumption of our model, that semantic structures derive from a growth process in which connections are introduced primarily between new nodes and existing nodes, predicts a causal relationship between the history of a network's growth and its ultimate pattern of connectivity. This relationship in turn motivates a unifying explanation for some previously observed behavioral phenomena of memory search, under the hypothesis that search processes exploit the structure of semantic networks in something like the way that state-of-the-art algorithms for web searching (Brin & Page, 1998) exploit the link structure of the web.

Most generally, our growth model predicts a correlation between the time at which a node first joins the network and the number of connections that it ultimately acquires. More precisely, at any given time, older nodes should possess more connections than younger nodes, and this effect should interact with any variation in utility (e.g., word frequency) that influences the probability of connecting new nodes to particular existing nodes. Figure 8 illustrates both the basic age effect and the nonlinear interaction with utility, using our undirected model of the word association data (model A) and utility distributed according to the log Kucera-Francis frequencies. (The directed network model shows similar results.) The age effect is strongest for nodes with highest utility because they acquire new connections at the highest rate.

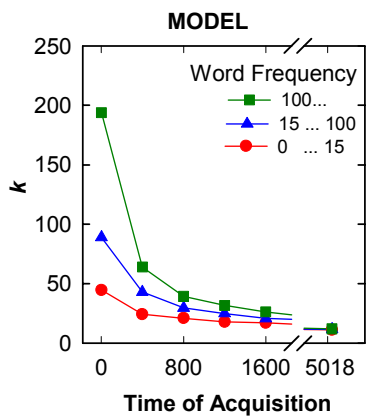

**Figure 8**. In the growing network models, the degree of a node decreases nonlinearly as a function of the time since it was first connected to the network. This age effect holds regardless of variations in node utility (e.g., based on word frequency) and is greatest for the highest utility nodes. Here the nodes are binned into three sets corresponding to low, medium and high frequency in the norms of Kucera and Francis (1967).

In order to test these predictions of the model, we consulted age-of-acquisition norms that are available for small sets of words. Gilhooly and Logie (1980) asked adults to estimate, using an arbitrary rating scale, the age at which they first learned particular words. These ratings were converted to scores between 100 to 700, with a score of 700 corresponding to a word acquired very late in life. We took the average score for each word as a crude measure of the time at which that word typically enters an individual's semantic network. We also consulted the age-of-acquisition norms of Morrison et al. (1997), who in a cross-sectional study estimated the age at which 75% of children could successfully name the object depicted by a picture. While these norms provide a more objective measurement for the age of acquisition, they were only available for a very small set of words.

Figure 9 shows the relationships between the number of connections that a word possesses, as measured in each of the three real semantic networks analyzed earlier, and its age-of-acquisition, as measured by both the adult rating and child picture-naming norms. We separated the words into three different frequency bins to show interactions between age-of-acquisition and word frequency. The data for all three networks are similar, but the word association network is most directly comparable to the model predictions shown in Figure 8 (because the model was simulated to have the same size and density as this network). For both norms, early-acquired words have more connections than late-acquired words. Also as predicted by the model, high-frequency words show higher connectivities, and the effect of age-of-acquisition on connectivity is most dramatic for high-frequency words.

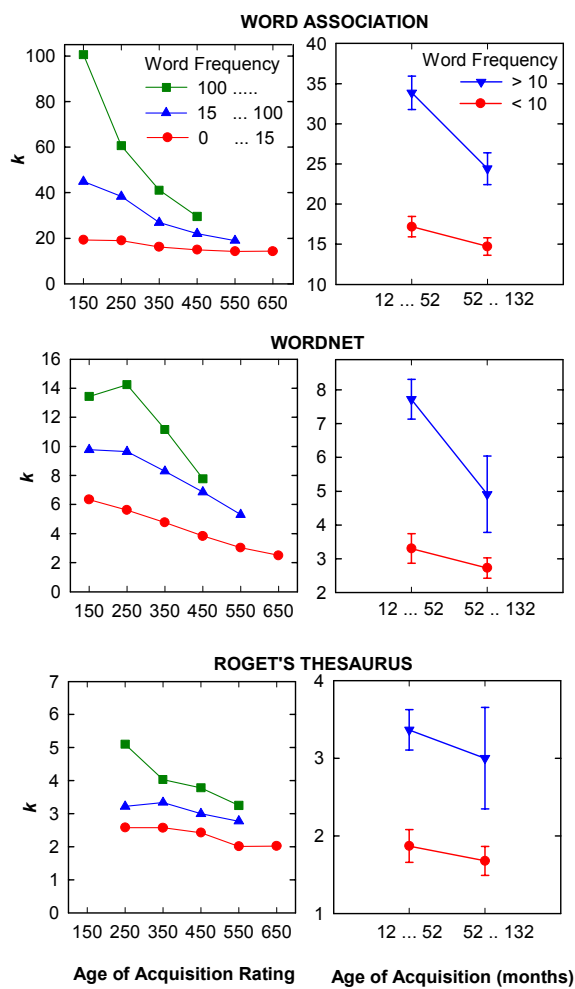

**Figure 9**. The relation between degree and age of acquisition as measured by adult ratings (left panels) and the average age at which children can name pictures (right panels). Right panels include standard error bars around the means.

Beyond the implication that semantic structure arises through some kind of temporally extended growth process, these relationships between connectivity, frequency, and age of acquisition do not in themselves place strong constraints on the mechanisms of semantic network formation. In contrast with the small-world and scale-free structures discussed earlier, many different growth models would make similar predictions here. More significant is the possibility that some model of semantic structure based on growth processes may offer an elegant unifying explanation for the behavioral effects of all of these variables.

Both high frequency and early age-of-acquisition have been shown to correlate with short reaction-time latencies in naming (e.g., Carroll & White, 1973) and lexical decision tasks (e.g., Turner, Valentine, & Ellis, 1998). While it has been suggested that age-of-acquisition affects mainly the speech output system (Ellis & Lambon Ralph, in press), age of acquisition has also been show to affect non-phonological tasks involving face recognition and semantic tasks, such as word association and semantic categorization (e.g., Brysbaert, Van Wijnedaele, & De Deyne, 2000). There has also been some debate on whether age-of-acquisition influences behavior independently of word frequency (e.g., Turner et al., 1998), or instead merely embodies cumulative frequency (Lewis, Gerhand, & Ellis, 2001), because high-frequency words are likely to be acquired earlier than low-frequency words.

In debating the causal status of variables such as frequency and age-of-acquisition, there are both functional ("why") questions and mechanistic ("how") questions. Functionally, it is clear that a bias towards high-frequency words or concepts would be useful in many cases, but it is not so clear what direct purpose an age-of-acquisition bias would serve. Mechanistically, there is good reason to doubt that either frequency or age-of-acquisition really are direct causes of people's behavior; it seems impossible for the history of learning to influence present behavior unless it has somehow left a structural trace in memory.

Our model of network growth suggests one possible structural substrate that could mediate the influences of both high frequency and early age-of-acquisition: the size of a node's neighborhood, or the number of connections it makes to other nodes, which we have shown correlates with both of these history factors. Mechanistically, such a bias could arise naturally if memory search is implemented by some kind of serial or parallel Markov process operating on a semantic network. Under many different probabilistic dynamics, a Markov search process would tend to find first those nodes with highest degrees (or in-degrees, for directed networks).

In addition to providing to providing one clear mechanism for how the history of a word's usage could affect present behavior, a bias to access words with high connectivity also has an intriguing functional basis. In Google, a state-of-the-art algorithm for web searching (Brin & Page, 1998), sites are ranked in part based on a measure of their centrality in the WWW. The centrality of a node reflects both its authority – the extent to which other sites point to it – as well as the probability that it will be encountered in random walks on the network. Google measures the centrality of a node according to its projection onto the principal eigenvector of the normalized web adjacency matrix. Essentially the same computation applied to a graph of feature dependencies, rather than WWW connections, has been used to assess the conceptual centrality of different object features (Sloman, Love, & Ahn, 1998). The (in-)degree of a node provides a simpler alternative measure of centrality, which typically correlates highly with eigenvector measures of centrality and thus authority as well. Just as a Google search orders websites based on their centrality, so might the cognitive search processes involved in word production or lexical retrieval be functionally biased towards accessing more central, highly connected nodes first, as a means to direct processing towards the most authoritative and causally deep concepts that could be employed in a particular situation.

As a preliminary investigation of the behavioral significance of degree centrality, we undertook a correlational analysis of word frequency, age-of-acquisition, and node degree factors in predicting reaction times for several psycholinguistic tasks. Specifically, we looked at naming and lexical decision in two databases; a new naming latency database by Spieler and Brand (personal communication) for 796 multi-syllabic words and a large lexical decision latency database from Balota, Cortese, and Pilotti (1999) for 2905 words. Node degrees were logarithmically transformed to avoid the extreme skew inherent in the degree distribution. Table 2 shows the correlations between latencies in naming and lexical decision tasks and the three factors of degree centrality, age of acquisition (referred to as AoA) and word frequency. The results confirm well-known findings that age of acquisition correlates positively with naming and lexical decision latencies and that word frequency (using the norms of Kucera and Francis) correlates negatively with the same latencies. In other words, high-frequency or early-acquired words are named faster and are quicker to be identified as words than low-frequency or late-acquired words. We also see that centrality is negatively correlated with these latencies: words that are semantically more central have quicker reaction times. All of these correlations are relatively weak, but most are statistically significant (those with $n>100$). When the effects of word frequency or age of acquisition are partialed out of the analysis, the correlations between centrality and reaction-time latencies become weaker but remain significant for the lexical decision task.

**Table 3**. Correlations between naming and lexical decision latencies and three potential causal factors: word frequency, age of acquisition and degree centrality (in the word association, WordNet, and Roget's networks).

|  | Naming | | Lexical Decision | |
| --- | --- | --- | --- | --- |
|  | R | n | R | n |
| Log( k ) - Word Association | -.330 * | 466 | -.463 * | 1676 |
| Log( k ) - Wordnet | -.298 * | 790 | -.464 * | 2665 |
| Log( k ) - Roget | -.164 * | 647 | -.253 * | 2343 |
| Log( word frequency ) | -.333 * | 713 | -.511 * | 2625 |
| AoA (rating) | .378 * | 199 | .551 * | 566 |
| AoA (picture naming) | .258 | 44 | .346 * | 137 |
| **After partialing out log( word frequency )** | | | | |
| Log( k ) - Word Association | -.194 * | 433 | -.258 * | 1634 |
| Log( k ) - Wordnet | -.171 * | 706 | -.274 * | 2503 |
| Log( k ) - Roget | -.110 * | 602 | -.136 * | 2243 |
| AoA (rating) | .337 * | 196 | .450 * | 546 |
| AoA (picture naming) | .208 | 39 | .239 * | 131 |
| **After partialing out AoA (picture naming)** | | | | |
| Log( k ) - Word Association | -.279 | 33 | -.414 * | 107 |
| Log( k ) - Wordnet | -.246 | 36 | -.394 * | 111 |
| Log( k ) - Roget | -.141 | 29 | -.195 * | 105 |
| Log( word frequency ) | -.280 | 34 | -.463 * | 109 |
| **After partialing out log( word frequency ) & AoA (picture naming)** | | | | |
| Log( k ) - Word Association | -.171 | 32 | -.234 * | 106 |
| Log( k ) - Wordnet | -.145 | 33 | -.242 * | 108 |
| Log( k ) - Roget | -.101 | 33 | -.104 | 104 |

Note: R=correlation; n=number of observations; * is placed next to significant correlations (p<.05)

For the most part these correlations are not surprising, given that we have already shown a correlation – and, in our growth model, a causal dependence – between degree of connectedness and the two history factors of age-of-acquisition and usage frequency. However, they suggest that in addition to perhaps being one mediator of learning history, degree centrality may also exert an independent effect on reaction times. This bolsters our hypothesis that memory search processes are dependent either functionally or mechanistically on the connectivity structure of semantic nets.

We are not suggesting that degree centrality is the only structural locus for learning history effects, nor that the causal relationships between these factors only point in the one direction captured by our growth model. In particular, our model explains why more frequent, early-acquired words are more densely connected, but it does not explain why certain words are acquired earlier or used more frequently than other words. There could well be a causal influence in the other direction, with age of acquisition and usage frequency dependent on the degree of connectivity in the semantic networks of competent adult speakers in a language community.

Our growing network model is also not the only model that attempts to capture the structural effects of age of acquisition and frequency. In several connectionist models (Ellis & Lambon Ralph, in press; Smith, Cottrell, & Anderson, 2001), it has been found that the final accuracy for a particular training pattern depends on both its frequency of presentation and its age of acquisition (manipulated by hand in Ellis & Lambon Ralph (in press), and estimated by Smith et al. (2001) to be the first time at which the training error drops below some threshold). The connectionist explanations depend upon distributed representations: early-learned items induce a distributed representation that later-learned items cannot easily change, and thus over time, the model loses the ability to encode new patterns effectively. Our growing network model offers an alternative view in which each word or concept is represented as a distinct entity, together with explicit connections to other entities. In contrast with the distributed representations account, this view has several distinctive features. Age-of-acquisition effects do not imply that recently learned words are encoded any less effectively than older ones, only that they are less richly connected to other items in memory. There is also a clear route for how age-of-acquisition could influence the dynamics of search behavior, via the structural feature of degree centrality that is both functionally and mechanistically relevant for effective network search procedures, and which seems to exert an independent effect on reaction times. It is not so clear in connectionist accounts how or why the magnitude of a pattern's training error should determine its accessibility or reaction time in a search process.

## Conclusion

We have found that several semantic networks constructed by quite different means all show similar large-scale structural features: high sparsity, very short average path-lengths, strong local clustering, and a power-law distribution for the number of semantic neighbors, indicating a hub-like structure for knowledge organization. These regularities are not predicted by conventional static models of semantic structure, but can be explained as the consequences of a simple network growth process based on semantic differentiation.

These statistical principles of large-scale semantic structure may be valuable beyond offering the potential to link the organization of mature semantic representations to the processes of language development or evolution. In the most general terms, robust quantitative empirical relationships typically provide some of the strongest constraints on computational models of cognition, and until now there have been few such laws for semantic structure. Any computational model that seeks to explain how semantic structures form and what shape they ultimately take will have to reckon with these results.

Models of semantic processing may also have to be sensitive to these structures because, in the words of one prominent network theorist, "structure always affects function." (Strogatz, 2001, p. 268). Since the seminal work of Collins and Quillian (1969), which explored the interaction between one simple kind of structure for semantic networks and its complementary processes, researchers have thought mainly in terms of general processes such as spreading activation operating on arbitrary structures. However, the finding of small-world and scale-free structures in semantic networks might cause us to rethink how search and retrieval could work in these networks. We have already suggested how retrieval processes in lexical decision and naming might be attuned to one aspect of semantic network connectivity – namely, node centrality – which leads as a natural consequence to effects of frequency and age-of-acquisition on reaction time. More generally, the processes involved in searching for relevant and useful knowledge in semantic networks might be adapted to their large-scale structure in any number of ways, and might be very different from search processes adapted to arbitrarily structured graphs (or other non-scale-free structures, such as inheritance hierarchies or high-dimensional Euclidean vector spaces).

The statistical picture we have described here could reflect two central aspects of language that often seem difficult to reconcile, except in the hands of the greatest writers. On the one hand, language provides an extremely flexible system of thinking. By putting just a few words together in a sentence, we can draw meaningful, often surprising, connections between practically any two concepts. Making a loose analogy to paths in a semantic network, this flexibility corresponds to the very short average path-lengths and network diameters observed in our analyses. Yet this flexibility does not come at the cost of general vagueness or disorder. On the contrary, linguistic meanings are organized into numerous coherent and distinct semantic domains, with many words having quite specific meanings in at most one or two of these domains. This organization is reflected in the high degrees of sparsity and neighborhood clustering that we observed.

What allows language to sustain both specificity and flexibility – in graph-theoretic terms, both sparse, clustered neighborhoods and short path lengths – is the existence of hubs: polysemous words or concepts that bridge multiple, and otherwise distinct, semantic regimes. In the theory of literature, these hubs correspond to symbols or serve as the foundations for metaphors. Indeed, the most expressive metaphors or symbols derive their power precisely from the conjunction of great specificity with great flexibility. They focus our attention on the familiar details of one subject, and then, without notice, connect that subject to a seemingly unrelated one, thereby showing us sides of both that we had not seen before. The greatest poets, writers, and speakers are masters of large-scale semantic structure. They intuitively understand the possibilities of small-world and scale-free structures much the same way that European painters since the Renaissance – long before contemporary vision scientists – came to understand the principles of projective geometry, color constancy, and shape-from-shading. As our formal knowledge of the large-scale organizational principles of meaning matures, it becomes increasingly possible to imagine a poetics in which those cognitive principles play an integral role.

Finally, it is always tempting to speculate on the possibilities for correspondence between semantic networks in the mind and neural networks in the brain. Our finding of scale-free structures in semantic nets cautions against drawing any simplistic analogies between nodes and neurons or connections and synapses. The neural network of the worm *C. Elegans* has been shown to have a small-world structure (Watts & Strogatz, 1998), but it is definitely not scale-free it its connectivity. The degree distribution falls off according to a very clear exponential law with a single characteristic scale (Amaral et al., 2000). Likewise, the connectivity of neurons within a cortical area may have some small-world features, due to the existence of excitatory connections within a local two-dimensional neighborhood and long-range inhibitory connections between neighborhoods (Kleinberg, 2000), but it is almost certainly not scale-free. There are typically only one or a few scales of connectivity in any cortical area, with each neuron making a number of connections that is fairly similar to other neurons of the same type (Kandel, Schwartz & Jessell, 1991). Any direct mapping from nodes in a semantic network onto single neurons or columns of neurons, and from the connections between nodes onto the synapses between those neurons, would therefore not preserve the essential scale-free structure of semantic relationships. More likely, the correspondence between semantic nets and neural nets takes the form of a functional mapping, implemented in the physiology of neural populations, rather than a structural mapping implemented in their microanatomy. When the time comes for a serious study of these mind-brain correspondences, the quantitative principles that we have presented here may provide one source of guidance in identifying the physiological basis of semantic representations.

# Notes

1. For reasons of space, we provide only heuristic definitions for some of these terms. See Watts (1999) for a more in-depth treatment of graph-theoretic concepts in connection to small-world networks.

2. Note that distances in an undirected graph always satisfy the three metric axioms of minimality, symmetry, and the triangle inequality (see Tversky, 1977), but that distances in an undirected graph do not in general satisfy the latter two.

3. We implemented Dijkstra's algorithm with Fibonacci heaps (Cormen, Leiserson, & Rivest, 1990) as an efficient means to find the shortest paths between a given node and all other nodes. (Matlab code for this algorithm is available from the first author.) For very large networks, it is often computationally infeasible to calculate the shortest paths between all pairs of nodes. In such cases, we can estimate $L$ and $D$ based on the shortest paths for a large sample of nodes (see note 6).

4. The second property holds simply because $P(k)$ in a random graph is a binomial distribution (Bollobás, 1985), and all binomial distributions have this shape. The first trend occurs because a lower value of $C$ – lower overlap in the neighborhoods of neighboring nodes – implies that on average, more distinct nodes can be reached in two steps from any given node. At the extreme value of $C = 0$, none of the nodes that can be reached by taking two steps from node $i$ is also a neighbor of $i$, and thus the number of nodes within a given distance of each node grows exponentially with distance. As $C$ increases, more of the paths radiating out from a node become redundant and more steps are required on average to connect any two nodes.

5. For WordNet, there were connections between word and meaning nodes, between word and word nodes, and between meaning and meaning nodes; these connections were rearranged separately when constructing the random graphs.

6. For the word associative networks, $L$ and $D$ were calculated on the basis of the path lengths between all word pairs in the large (strongly) connected component. For the much larger networks of WordNet and Roget's thesaurus, $L$ and $D$ were based on the path lengths between all pairs of a sample of 10,000 words from the large connected component (see note 3).

7. Zipf (1965) plotted the number of meanings of a word versus its rank of its word frequency in log-log coordinates and observed a slope b=.466. Adamic (2000) provides some simple analytic tools by which the slope b=.466 in this Zipf plot can be converted to $\gamma$=3.15, the slope of the corresponding probability distribution in log-log coordinates.

8. Both the path-lengths and the clustering coefficients of the LSA networks show a decreasing trend as the dimensionality $d$ is increased. It is possible that at dimensionalities higher than 400, these statistics will come closer to the values observed for the word association network. We were not able to investigate this possibility, as only 400 LSA dimensions were available to us. However, it is unlikely that increasing the dimensionality would threaten our main argument, because the lack of scale-free degree distributions is the primary feature distinguishing the LSA networks from naturally occurring semantic networks and our growing network models. Based on Figure 7, it seems doubtful that these distributions would match significantly better at higher dimensionalities (unless perhaps $d$ was increased far beyond the typical value of 300).

# Acknowledgments

We are grateful to Tom Landauer and Darrell Laham for providing us with the TASA corpus. We also thank Eve Clark, Tom Griffiths, Doug Nelson, and Vin de Silva for discussions that helped shape this research. Portions of this work were previously described in a shorter unpublished manuscript (Steyvers & Tenenbaum, 2001) that was circulated on the internet:
(http://www-psych.stanford.edu/~msteyver/papers/smallworlds_cogsci.pdf).